\documentclass[acmlarge]{acmart}
\AtBeginDocument{%
  \providecommand\BibTeX{{%
    \normalfont B\kern-0.5em{\scshape i\kern-0.25em b}\kern-0.8em\TeX}}}

\usepackage{algorithm}
\usepackage{algpseudocode} 
\usepackage{multirow}
\usepackage{wrapfig}
\usepackage{array}
\usepackage{graphics}
\usepackage{amsmath}
\usepackage{nccmath}
\usepackage{mathtools}
\usepackage{caption}
\newcolumntype{M}[1]{>{\centering\arraybackslash}m{#1}}

\usepackage[toc,page]{appendix}

\DeclarePairedDelimiter\floor{\lfloor}{\rfloor}
\begin{document}

%%
%% The "title" command has an optional parameter,
%% allowing the author to define a "short title" to be used in page headers.
\title{Diagnostic Assessment Generation via Combinatorial Search}

%%
%% The "author" command and its associated commands are used to define
%% the authors and their affiliations.
%% Of note is the shared affiliation of the first two authors, and the
%% "authornote" and "authornotemark" commands
%% used to denote shared contribution to the research.

\author{Daehan Kim}
\authornote{Authors contributed equally to this research.}
\email{daehan_kim@tmax.co.kr}
\affiliation{%
  \institution{TmaxEdu Inc.}
  \city{Seongnam}
  \state{Gyeonggi}
  \country{South Korea}
}

\author{Hyeonseong Choi}
\authornotemark[1]
\authornote{This research was conducted while at TmaxEdu Inc.}
\email{hschoi@smartradarsystem.com}
\affiliation{%
  \institution{Smart Radar System Inc.}
  \city{Seongnam}
  \state{Gyeonggi}
  \country{South Korea}
}

\author{Guik Jung}
\authornotemark[1]
\email{guik_jung@tmax.co.kr}
\affiliation{%
  \institution{TmaxEdu Inc.}
  \city{Seongnam}
  \state{Gyeonggi}
  \country{South Korea}
}

%%
%% By default, the full list of authors will be used in the page
%% headers. Often, this list is too long, and will overlap
%% other information printed in the page headers. This command allows
%% the author to define a more concise list
%% of authors' names for this purpose.
% \renewcommand{\shortauthors}{Anonymous Author(s)}

%%
%% The abstract is a short summary of the work to be presented in the
%% article.
\begin{abstract}
 Initial assessment tests are crucial in capturing learner knowledge states in a consistent manner. Aside from crafting questions itself, putting together relevant problems to form a question sheet is also a time-consuming process. In this work, we present a generic formulation of question assembly and a genetic algorithm based method that can generate assessment tests from raw problem-solving history. First, we estimate the learner-question knowledge matrix (snapshot). Each matrix element stands for the probability that a learner correctly answers a specific question. We formulate the task as a combinatorial search over this snapshot. To ensure representative and discriminative diagnostic tests, questions are selected (1) that has a low root mean squared error against the whole question pool and (2) high standard deviation among learner performances. Experimental results show that the proposed method outperforms greedy and random baseline by a large margin in one private dataset and four public datasets. We also performed qualitative analysis on the generated assessment test for 9th graders, which enjoys good problem scatterness across the whole 9th grader curriculum and decent difficulty level distribution.
 
 %% relavant -> relevant 수정 - 최현성
 %% an combinatorial -> a combinatorial 수정 - 최현성
 %% four public dataset- > four public datasets 수정 - 최현성
 %% individual -> individual(chromosome) 수정

\end{abstract}

%%
%% The code below is generated by the tool at http://dl.acm.org/ccs.cfm.
%% Please copy and paste the code instead of the example below.
%%
% \begin{CCSXML}
% <ccs2012>
%  <concept>
%   <concept_id>10010520.10010553.10010562</concept_id>
%   <concept_desc>Computer systems organization~Embedded systems</concept_desc>
%   <concept_significance>500</concept_significance>
%  </concept>
%  <concept>
%   <concept_id>10010520.10010575.10010755</concept_id>
%   <concept_desc>Computer systems organization~Redundancy</concept_desc>
%   <concept_significance>300</concept_significance>
%  </concept>
%  <concept>
%   <concept_id>10010520.10010553.10010554</concept_id>
%   <concept_desc>Computer systems organization~Robotics</concept_desc>
%   <concept_significance>100</concept_significance>
%  </concept>
%  <concept>
%   <concept_id>10003033.10003083.10003095</concept_id>
%   <concept_desc>Networks~Network reliability</concept_desc>
%   <concept_significance>100</concept_significance>
%  </concept>
% </ccs2012>
% \end{CCSXML}

% \ccsdesc[500]{Computer systems organization~Embedded systems}
% \ccsdesc[300]{Computer systems organization~Redundancy}
% \ccsdesc{Computer systems organization~Robotics}
% \ccsdesc[100]{Networks~Network reliability}

%%
%% Keywords. The author(s) should pick words that accurately describe
%% the work being presented. Separate the keywords with commas.
\keywords{Diagnostic Assessment Generation, Knowledge Tracing, Combinatorial Search, Genetic Algorithm, Knowledge Snapshot}

%%
%% This command processes the author and affiliation and title
%% information and builds the first part of the formatted document.
\maketitle

\section{Introduction}

Designing good diagnostic tests are crucial in assessing learner knowledge state. Since learners cannot be asked to solve all questions in a domain just to measure his/her performance, small number of selected questions are required and deployed in many standardized tests. For instance, TOEIC measures business English communication skill in a score of 0-1000, with 100 reading comprehension and 100 listening comprehension problems. However, manually designing such question sheets is costly and time-consuming. Our research question arises here : \textbf{Can we select representative diagnostic questions when learners' problem solving history is given?} 

We first observed there are no standard problem formulation for generating diagnostic tests. In this work, we define this task as a combinatorial search problem, where we search for a subset of problems that minimizes the learners performance gap between whole problems and selected problems, and maximize discrimination among learners. More specifically, learner performance gap is computed by Root Mean Squared Error of average score on selected questions against the whole set average score across all learners. Also, discrimination among learners can be defined using standard deviation of average score on selected questions across all learners.

We then present a pipeline for selecting such a subset of questions that optimize above criterion. In the framework, we first generate snapshot of learners, which is learner-question matrix $Q$ containing probabilities $Q_{ij}$ that learner j solves problem i, using generic Knowledge Tracing models. To scale to number of learners, learners are first randomly selected to reduce snapshot size. By this process, computing criteria becomes efficient with smaller snapshot size. Finally, we introduce a genetic algorithm based model that can find optimal subset of question pool with affordable computation cost of $\mathcal{O}(N_gN_p)$, where $N_g$ is the number of generation, and $N_p$ size of population.

Comprehensive experiments show that our proposed method outperforms both random and greedy baseline algorithm across four public and one private dataset. We also present qualitative analysis on generated question sheets to show effectiveness of our method.

%% both expert designed rule and greedy baseline algorithm -> both random rule and greedy baseline algorithm 으로 수정해야 하지 않나요? - 최현성

% These days, Knowledge Tracing(KT) has become a backbone for computer-aided education across many online learning platforms. Basic idea of KT is that learner knowledge state, which is represented by the probability of solving specific question, can be modeled using sequence of learner-problem interactions either by statistical model or deep neural architectures. Most KT models are basically user-agnostic : they are trained with many different learners' interactions and are to capture commonly observed knowledge fluctuation pattern. They require some initial learning records at inference phase to specify learner states well. [needs cites] We speculate that providing such a good set of initial questions can improve trained knowledge model inference. 

Our contributions are two-folds.
\begin{itemize}
    \item We formulate diagnostic assessment generation as a combinatorial search problem where a set of problems that maximize similarity to the entire set and discrepancy among learner performances, without any assumption on model architecture. 
    \item We propose a genetic algorithm based method that searches for diagnostic question sheet in a given question pool. Experimental results show our method outperforms two basic baselines and is able to generate decent quality diagnostic question sheets. 
\end{itemize}

\begin{figure}[H]
\centering
\includegraphics[width=1\linewidth]{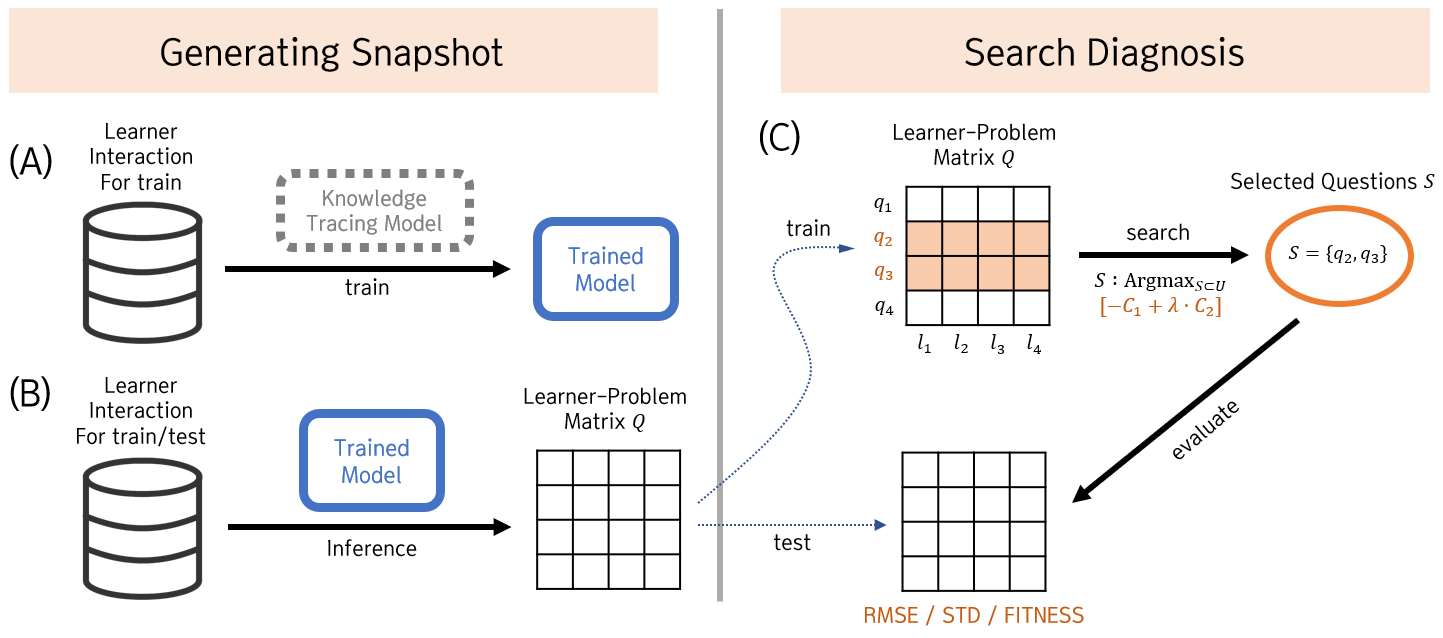}
  \caption{(A) Knowledge Tracing models are trained using sequential learner-question response dataset. Trained model can output predicted learner's solving probability for all questions when fed with response history. (B) Learner-Problem matrix (Snapshot) is generated using trained knowledge tracing model and interaction history. (C) Search Algorithms are applied to find a question subset that maximizes criterion $C(S)$. For fair evaluation, criteria are computed over unseen learners. }
\end{figure}

\section{Problem Formulation}

We formulate diagnosis generation task as a combinatorial search problem. Given a set of questions $U = \{q_1,q_2,...,q_{|U|}\}$ and search size $K$, we aim to finding a subset $S \subset U$ such that $|S| = K$ that maximizes the predefined criterion $C:S \rightarrow \mathbb{R}$. For brevity of further description, let us define some notations here. We denote the set of all learners as $L=\{ l_1, l_2, ..., l_{|L|} \}$ and a set of all subsets of $U$ as $2^U$. For construction of the criterion, we consider following two factors:

\begin{itemize}
    \item $C_1$ : Discrepancy between holistic score and subset score 
    \item $C_2$ : Discrimination among learners
\end{itemize}

The criterion $C_1:2^U \times 2^U \rightarrow \mathcal{R}$ aims to measuring the difference between learners' score when all the problems $U$ are considered and those when only selected problems $S \subset U$ are considered. More specifically, when a learner $l$ is given, we have probability values $p_{ql}$ of each problem $q$ answered correctly. We get average values of two sets $\{ p_{ql} : q \in U \text{ and } l \in L \}$ and $\{ p_{ql} : q \in S \text{ and } l \in L \}$, each denoted as $\mu_{Ul}$ \text{ and } $\mu_{Sl}$. Then $C_1$ can be described as 
% \begin{fleqn}[\parindent]
    % \begin{equation*}
    $$C_1(U, S) = \sqrt{\frac{1}{|L|}\sum_{l \in L} (\mu_{Ul} - \mu_{Sl})^2}$$
    % \end{equation*}
% \end{fleqn}

The criterion $C_2:2^L \times 2^U \rightarrow \mathcal{R}$ quantifies how well selected subset $S \subset U$ discriminates learners' performances. Specifically, when a set of learners $L$ and subset $S$ is specified, we can define each learner's performance as $\mu_{Sl}$, the average value of $\{ p_{ql} : q \in S \text{ and learner} = l \}$. Then we can define $C_2(L, S) = \text{STD}(\{\mu_{Sl} : l \in L \})$, where $\text{STD}(\cdot)$ means standard deviation of numbers in the set. 

%% how we set lambda?
The final criterion is weighted sum of two criterion $C_1$ and $C_2$ and defined as follow: 

$$ C(S) = -C_1(U,S) + \lambda \cdot C_2(L,S) $$

For $C(S)$ to be maximized, $C_1$ needs to be smaller and $C_2$ larger. $\lambda$ is a mixing coefficient that determines the relative importance of second criterion against first criterion. Since two crieterions differ in scale, we set $\lambda \simeq \frac{\overline{C_1}}{\overline{C_2}}$, making both criterion have similar importance. $\overline{C_1}$ and $\overline{C_2}$ are averages of $C_1(U, S)$ and $C_2(L,S)$ with 10,000 of random sampling for $S$, respectively. By this choice of lambda, one can see whether a solution is better than random guess by comparing its fitness to zero.

\section{Methods}

We first generate learners' snapshot using trained knowledge tracing model and applied genetic algorithm to find a subset $S$ of questions that maximizes the criterion $C(S)$. 

\subsection{Knowledge Tracing Model}

Knowledge Tracing(KT) has become a de-facto standard to track learner knowledge state at a specific time frame. After DKT emerged in the field, most neural-network-based KT model accepts inputs of learner's problem solving history. For a single learner, it can be represented by vector $X = [x_{1}, x_{2}, ..., x_{T}]$, where each $x_{t}$ is one-hot embedding for question-response pair at time $t$ and $[ \cdot ]$ means vector concatenation. KT model then outputs probability vector $\hat{Y} \in \mathcal{R}^{|U|}$ that the learner would correctly answer each question at time $T$ in question pool $U$.   
We utilized Neural Pedagogical Agent(NPA) \cite{lee2019creating} as a base knowledge tracing model in this work.
NPA sucessfully deploys bi-LSTM to model sequential nature of knowledge and attention mechanism to focus on most influential interaction in the learning history. 
%% interaction은 뭐고 / interaction의 시퀀스는 LSTM에 전해지고 attention의 k,q,v는 뭔지, 문제를 풀 확률은 어떻게 계산되는지
%% What is input and output of NPA?
More specifically, we embedded question-response pair and difficulty level into $d$-dimensional vectors and used their addition as inputs for bi-LSTM. 
$$ E_{t} = \text{Embedding}(X_{t}) \in \mathcal{R}^d_v$$
$$ R_{t} = D_{t} \oplus E_{t} \in \mathcal{R}^d_v$$
where $\oplus$ is element-wise sum and $D_t$ is difficulty level embedding.
$$ v_t = \text{bi-LSTM}(R_{<t}) \in \mathcal{R}^d_v$$

Additive attention is computed with question $q \in \mathcal{R}^{d_q}$ as a query and $v_t$ as key and value. Learner embedding at timestep $t$ is as follow.

$$ \alpha_t = \text{attention}(v_t, q) \in \mathcal{R}$$

where $\text{attention}(v_t, q) = \text{softmax}([\text{tanh}(W_v \cdot v_{t'} + W_q \cdot q)]_{t'<T})_t$ and  $W_{v} \in \mathcal{R}^{1 \times d_v}, W_q \in \mathcal{R}^{1 \times d_q}$.
$$ l_{t} = \sum_{t'<t} {\alpha_{t'} \cdot v_{t'}} \in \mathcal{R}^d$$

Finally, correctness probability for question $q  \in \mathcal{R}^{d_q}$ at timestep $t$ is computed.

$$ \hat{P}_{tq} = f(l_{t}, q) $$

Training is performed with standard binary cross-entropy objective, using Adam optimizer\cite{kingma2014adam}.
%% 어떻게 난이도 피쳐가 만들어지는가?
As stated above, we used difficulty level embedding along with question responses. Difficulty level embedding is defined by question's correct ratio in training dataset into 3 categories : easy, normal, hard. Thresholds for each category are determined according to the correct ratio distribution of each dataset, in a way that meets the ratio easy : normal : hard = 3 : 4 : 3. More details on the training of NPA model are reported in the Appendix \ref{appendix:ktmodel}.

\subsection{Learner Performance Snapshot}

%% snapshot meaning
Performance of a learner changes whenever he/she proceeds to solve exercises and do other learning activities. Thus we call the question-learner matrix $Q \in \mathbb{R}^{|U| \times |L|}$ learner performance snapshot, whose element $Q_{ij}$ denotes probability that learner $l_j$ correctly answers question $q_i$.

%% we simulate Q using knowledge tracing model
Since knowledge tracing models probabilities $Q_{\cdot i} \in \mathbb{R}^{|U|}$ when a learner $l_i$ is specified, we can predict $Q$ matrix using trained knowledge tracing models of any kind. In our work, we generated snapshot using NPA(Neural Pedagogical Agent, \cite{lee2019creating}) since its Area Under Curve(AUC) was the highest in AIHUBmath dataset. We trained the model using the training split and applied it to every learner in whole dataset to generate a snapshot.  

%% I-Scream -> AIHub-math 수정 (최현성)

\begin{wrapfigure}{r}{0.49\textwidth}
  \vspace{-20pt}
  \begin{center}
    \includegraphics[width=0.49\textwidth]{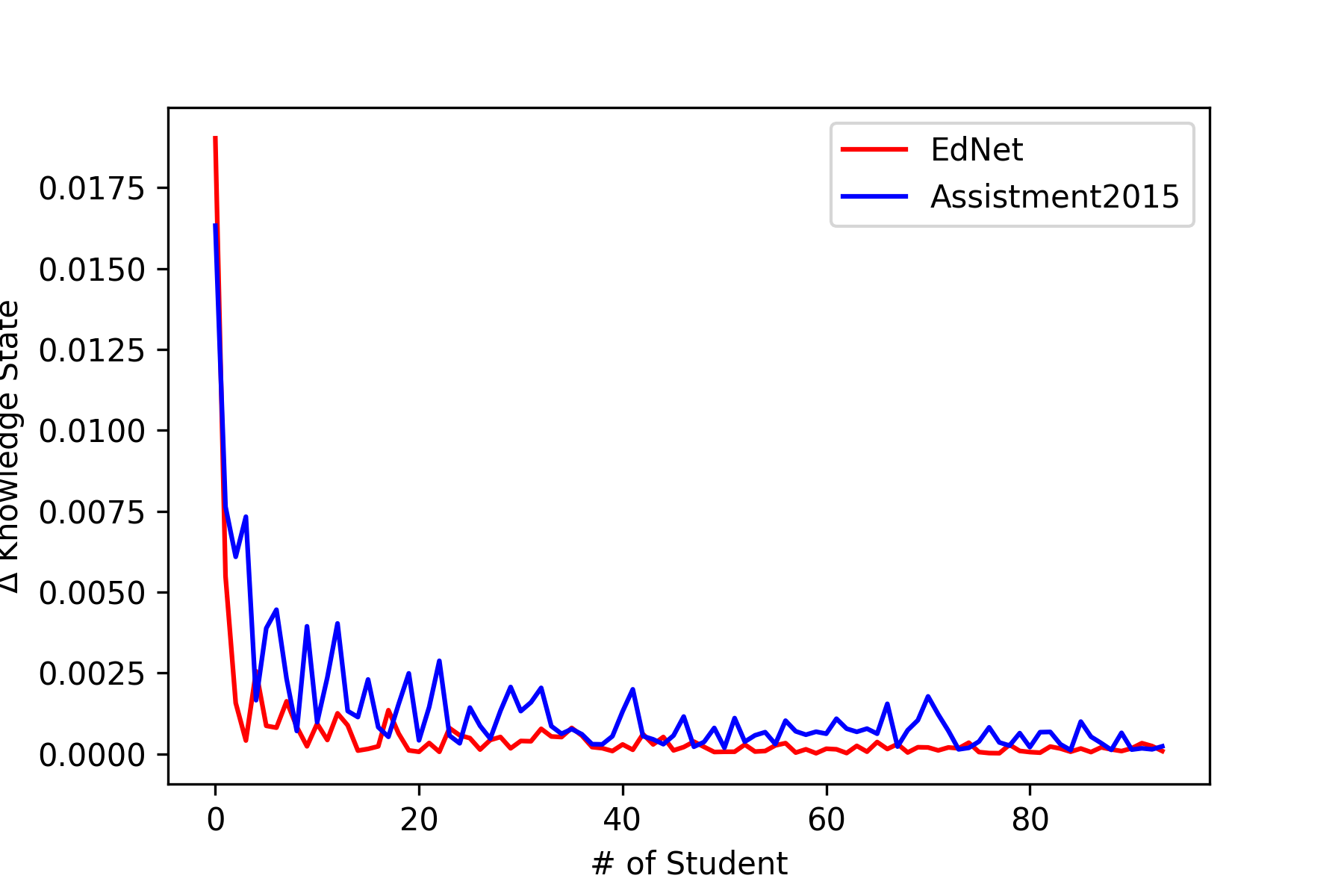}
  \end{center}
  \vspace{-10pt}
  \caption{Learners absolute performance change when new learners are added in each snapshot. This is generated using training split of each dataset and number of learners has scale of 100.}
\vspace{-20pt}
\label{fig:problem_number}
\end{wrapfigure}

%% 학생이 많거나 문제가 많을 경우 어떻게 scalable하게 스냅샷을 이용해 서치할수 있는가?
%% 학생이 많은 경우 쳐내면 됨. 
A set of learner scores $Q_{q} = \{ Q_{ql} : l \in L \}$ can be regarded as samples from performance distribution of question $q$. As law of large numbers states, the average of $Q_q$ approaches true learner score $\mu_q$ as the number of learners goes to infinity. We searched for practical large number $N$ over each dataset such that change of learner performance becomes near zero. Fig \ref{fig:problem_number} shows average performance change when new learners are added in a snapshot. EdNet and ASSISTment2015 have 784,309 and 19,840 students respectively and from the plot, 2 to 8k learners seem sufficient for estimating global average performance. We selected 6k for AIHUBmath and ASSISTment2009 dataset. One can also apply this strategy for each question and find the number of learners that stablize all problem's average performance.

%% model modification to advance AUC - AUC 증가를 위해 했다고 하기에는 논거가 부적절함. 애초에 multi uk랑 single uk는 비교 불가. 다만 난이도 반영은 성능 향상이 실제로 있었기 때문에 반영 가능. - 이 부분은 컨트리뷰션이 아니라 사족이다. 이상한 얘기가 될 수 있다.
% Since the optimization process is closely related to reliability of knowledge tracing model, we attached two minor modifications to the model : multi-concept handling and difficulty feature. Multi-concept handling enables the model to compute a probability of a question that has more than one knowledge concepts in an integrated manner. If a question has several knowledge concepts, its embedding is simply defined as the vector average of embeddings of its knowledge concepts. If a question contains only one knowledge concept, its embedding coincides with that of its knowledge concept. %% effects : does it increase performance than just averaging after probability is computed?
% Difficulty feature is easily derived from learner interactions by computing each question's correct ratio and categorizing questions into three difficulty domains(high, middle, low). Specifically, we sort each question by its correct ratio and split them by high:middle:low $= 3:4:3$ proportion. For ASSIST2015 dataset that does not provide question-level interactions, we employed NPA model without difficulty feature.

\subsection{Genetic Algorithm}
%% what is genetic algorithm? - fitness is defined above
%% how we defined cross-over / mutation?
Genetic algorithm is an algorithm that obtain superior individuals by evolving them over generations. The superiority of each individual $S$ can be assessed by \textbf{fitness} $C(S)$ mentioned above. Individuals stand for diagnostic tests, and a gene for a question. An individual $S$ is represented by a set of genes.
$$S = \{q_1, q_2, q_3, ..., q_k\} : q_i \in U$$
Likewise population is defined as a set of individuals. After these notations being set, we describe genetic algorithm as follow:

\begin{enumerate}
\item \textbf{Production:} Create a population of P individuals which are randomly selected subset of size $k$. 
\item \textbf{Selection:} Randomly pick 10\% of the individuals among the population and choose one individual with the highest fitness. Repeat process P times to produce P individuals at the end, which are regarded as the population of this generation.
\item \textbf{Crossover:} Pair up odd numbered individual with even numbered one, generating $\floor{N/2}$ pairs for $N$ individuals. Uniformly selected random index decides split position of two individuals and those splits are interchanged with probability $p_c$. Overlapping genes are replaced randomly to ensure the number of unique genes.

%% Overlapping genes are ignored to ensure the number of unique genes -> Overlapping genes change to other new genes to ensure a certain number of genes. 으로 변경 제안합니다. 실제로 무시되는게 아니라 겹치는 gene이 있을 경우, 새로운 다른 유전자로 변경(변이)됩니다. (최현성)

\item \textbf{Mutation:} Select individuals with probability $p_{m_1}$, where each gene is to be replaced randomly with probability $p_{m_2}$. Mutation process does \emph{not} allow genes that already exist in the individual.

%% by p_m1 -> with probability p_m1 수정. p_c, p_m2 부분과 통일 (최현성)

\end{enumerate}
Repeat (2) to (4) by the number of generations $N_g$ and obtain the individual with the highest fitness.
We implemented the genetic algorithm using DEAP library of Python. For more detail on algorithm, see Appendix \ref{appendix:pseudo}.

% \begin{wrapfigure}{r}{0.49\textwidth}
%   \vspace{-20pt}
%   \begin{center}
%     \includegraphics[width=0.49\textwidth]{Figures/difficulty.png}
%   \end{center}
%   \vspace{-10pt}
%   \caption{Correct Ratio distribution for EdNet dataset. Most questions are fairly easy. }
% \vspace{-30pt}
% \label{fig:difficulty_plot}
% \end{wrapfigure}

% \subsubsection{Problems}

% Problems are main components of snapshot and can be represented by their score distribution. To preserve question pool's representation ability, we sampled questions by 2-step procedure. First, we project the representation of each question $i$, which is performance vector $Q_i^{train}$ for training split learners, into low-dimensional space. Any unsupervised projection algorithm such as PCA or t-SNE can be used in the process. K-means clustering then is applied to cluster projected questions, where we sample questions that are close to cluster centroids. Number of samples per cluster is set proportionally by the sample size of each cluster. We applied this method to EdNet dataset, which has 13,169 questions, to sample 2,000 questions. 

\begin{table}
  
  \label{tab:freq}
\begin{tabular}{cccccc}
\toprule
\multirow{2}{*}{}           & \multicolumn{2}{c}{ASSISTment} & \multirow{2}{*}{EdNet} & \multirow{2}{*}{AIHUBmath} & \multirow{2}{*}{Simulated-5} \\ \cline{2-3}
                            & 2009           & 2015          &                        &                           &                              \\ \midrule
\# of Students              & 4,163          & 19,840        & 784,309                & 4,673                     & 6,000                        \\
\# of Questions             & 26,688         & 100           & 13,169                 & 2,887                     & 50                           \\
\# of Interaction           & 283,105        & 683,801       & 95,274,496             & 707,450                   & 300,000                      \\
Avg Interaction per Student & 68             & 34            & 121                    & 154                       & 50                           \\
Domain                      & Mathematics    & Mathematics   & English                & Mathematics               & Synthetic                            \\\bottomrule
\end{tabular}
    \vspace{10pt}
  \caption{Dataset Statistics}
  \vspace{-20pt}
\end{table}

\section{Datasets}

We describe 4 publicly available datasets (ASSISTment2009, ASSISTment2015, EdNet, Simulated-5) and a private dataset (AIHUBmath) in terms of their construction and statistics. 

%% I-Scream -> AIHub-math 수정 (최현성)

\subsection{AIHUBmath}

AIHUBmath Dataset is 7-9 graders' math question solving logs gathered from I-scream Edu online learning platform. Questions are bundled in question sheets, where questions in a sheet are all related to a specific knowledge concept. Questions are of multiple-choice and subjective types and can be interactively solved inside the platform. As a result, we gathered 707,450 problem solving interactions from 4,673 students. The average number of interactions per student is 153.84.

\subsection{ASSISTment}

ASSISTment dataset\cite{selent2016assistments} was gathered from a 8-graders' math learning platform, ASSISTment that gives learners consecutive scaffolding questions in case of first wrong response. Scaffolding questions consists of ones related to the main question and help learners understand the underlying solving process. We deployed 2 different versions, ASSISTment2009, 2015. They were collected for one year in 2009 and 2015 respectively from the same education platform.

%% ednet
\subsection{EdNet}
EdNet\cite{choi2020ednet} is a large collection of various learning activities from TOEIC learning app, \emph{Santa}. KT1-4 sets are provided according to their granularity, where KT1 only covers basic problem solving interactions and durations whereas KT4 contains more meta-activities such as purchasing learning packages or pausing an audio clip at specific cursor time. It is constructed from activity logs generated in \emph{Santa}, from 2017 to 2020. Due to the app's popularity, the dataset(KT1) contains 95,293,926 student interactions from 784,309 students with 121.5 average interactions per student. To the best of our knowledge, it is the largest public IES dataset.

\subsection {Simulated-5}

Simulated-5\cite{piech2015deep} is a synthetic dataset that simulates learners based on Item-Resonse theory. Each problem in the dataset has a concept id $i \in \{0,1,2,3,4\}$ and difficulty $\alpha \sim \mathcal{N}(0,1)$. Each learner has concept skill $\beta \in \mathbf{R}^5$ with each element following $\mathcal{N}(0,1)$. With slip probability $c$, probability that learner $l$ solves problem $q$ correctly is calculated by $P(X_{ql} = 1 | \alpha, \beta, c) = c + \frac{1-c}{1+e^{\alpha - \beta}}$. As \cite{piech2015deep} suggested, each concept skill increases with constant factor $\delta_q$ when students answers question $q$ correctly. This $\delta_q$ is sampled from $\mathcal{N}(0.4,0.05^2)$ distribution. In this dataset only, each learner is required to solve all 50 questions sequentially in increasing order. We adopted this synthetic dataset to compare true snapshot and predicted snapshot and for learner simulation purposes.

%% dataset statistics table : which domain?, interaction #, student #, avg interaction per student, gathering period, knowledge concept #, whether-it-is-multi-uk and so on

\begin{table}[]
\centering
\resizebox{\columnwidth}{!}{%
\begin{tabular}{ccccc|ccc|ccc}
\toprule
\multicolumn{2}{c}{\multirow{2}{*}{}}                                           & \multicolumn{3}{c|}{Random}     & \multicolumn{3}{c|}{Greedy}                              & \multicolumn{3}{c}{GA}                 \\ \cline{3-11} 
\multicolumn{2}{c}{}                             & RMSE     & Std.     & Fitness   & RMSE     & Std.                               & Fitness  & RMSE     & Std.     & Fitness           \\ \midrule
\multicolumn{1}{c|}{\multirow{2}{*}{ASSISTment}} & \multicolumn{1}{c|}{2009}    & 0.045140 & 0.166305 & -0.008470 & 0.011207 & 0.168501                           & 0.025947 & 0.004099 & 0.170038 & \textbf{0.033395}$^\ast$ \\
\multicolumn{1}{c|}{}                            & \multicolumn{1}{c|}{2015}    & 0.057359 & 0.109682 & -0.011928 & 0.016169 & 0.113232                           & 0.030731 & 0.012176 & 0.111334 & \textbf{0.033938}$^\ast$ \\ \hline
\multicolumn{1}{c|}{\multirow{3}{*}{AIHUBmath}}   & \multicolumn{1}{c|}{Grade 7} & 0.031128 & 0.222232 & 0.010518  & 0.007564 & 0.223940                           & 0.034401 & 0.005142 & 0.223840 & \textbf{0.036804}$^\ast$ \\
\multicolumn{1}{c|}{}                            & \multicolumn{1}{c|}{Grade 8} & 0.039918 & 0.227920 & -0.002972 & 0.005828 & 0.231175                           & 0.031644 & 0.004745 & 0.230005 & \textbf{0.032538}$^\ast$ \\
\multicolumn{1}{c|}{}                            & \multicolumn{1}{c|}{Grade 9} & 0.039199 & 0.248258 & -0.003648 & 0.005909 & 0.243549                           & 0.028966 & 0.004623 & 0.242634 & \textbf{0.030122}$^\ast$ \\ \hline
\multicolumn{2}{c|}{Simulated-5}                                                & 0.042539 & 0.058993 & -0.008347 & 0.013894 & 0.070500 & 0.026967 & 0.010530 & 0.066921 & \textbf{0.028258}$^\ast$ \\ \hline
\multicolumn{2}{c|}{EdNet}                                                      & 0.027165 & 0.151822 & 0.008573  & 0.004267 & 0.146537                           & 0.030227 & 0.003801 & 0.146481 & \textbf{0.030680}$^\ast$ \\ \bottomrule
\end{tabular}%
}
\caption{\label{tab:ExpResult} Results are from held-out test split and are averaged over 10 different runs. $^\ast$ means statistical significance compared to second-best result with $p$ value of 0.01 . }
\vspace{-10mm}
\end{table}

\section{Results and Analysis}

%% fitness / rmse / std for expert system, random system, ours
%% our system pervails even with less concepts.
\subsection{Baselines}

We built a $K$-question diagnosis sheet according to the following baselines. We did not use curriculum information in both our method and baselines. We focused on the effectiveness of learning history for the construction of diagnostic assessments. Inspecting on effects of curriculum information would be the scope of the future work.

\begin{itemize}
    % \item Heuristic : 2 problems are randomly sampled each from predefined 5 dominant category - Geometry, Analysis, Probability/Statistics, Algebra, Equations. All questions were manually labeled into each category by domain experts. Also, generated question sheets should be composed of 2 easy, 6 normal, 2 difficult questions. This policy is applied only to I-Scream dataset.
    \item Random : 10 non-overlapping questions are randomly sampled from the question pool. By this baseline, we can compare how good each algorithm is in each criterion $C_1$ and $C_2$. According to one's choice for snapshot generation method (especially snapshot scale) and type of dataset, proposed criteria can vary significantly. 
    \item Greedy : $K$ problems are sampled successively from the question pool, wherein newly added question is not overlapping with already sampled subset and maximize $C(S)$ greedily in each step. This algorithm iterates $K$ times over the question pool of size $N$, rendering its time complexity $\mathcal{O}(KN)$.
\end{itemize}

\subsection{Results}

Note that each algorithm uses no prior knowledge about curriculum. We compared Random, Greedy and Genetic Algorithm in terms of RMSE($C_1$), Std($C_2$) and Fitness($C$). Results on random baseline implies that it is fairly easy to select samples with good standard deviation but not easy with good RMSE, rendering RMSE more important than standard deviation.

\subsection{Analysis}

\subsubsection {Qualitative Analysis on Generated Diagnostic Tests}

Table \ref{tab:QualAnalysis} shows details on the diagnosis question sheet generated by genetic algorithm. Without any prior knowledge on curriculum structure or question difficulty, generated diagnostic test enjoys good question scatterness across all 9-grader's curriculum and decent difficulty distribution. For a detail, there are 5 questions related to second-order expressions(1318, 2180, 389, 1563, 278), which is the core of 9th grade Algebra. Also, it includes many important geometry concepts such as chords and tangent line. For those who want further inspection, we reported 7 and 8-grader's generated question sheets in the Appendix \ref{appendix:pseudo}.

% \subsubsection{Relationship between RMSE and Std}

% We plotted RMSE and standard deviation of randomly sampled $S$ from each snapshot.  They showed Pearson correlation of $0.1$ and Spearman correlation of $0.2$, which is quite independent of each other. This result justifies the selection of criterion $C_1$ and $C_2$.

\begin{wrapfigure}{r}{0.49\textwidth}
  \vspace{-20pt}
  \begin{center}
    \includegraphics[width=0.49\textwidth]{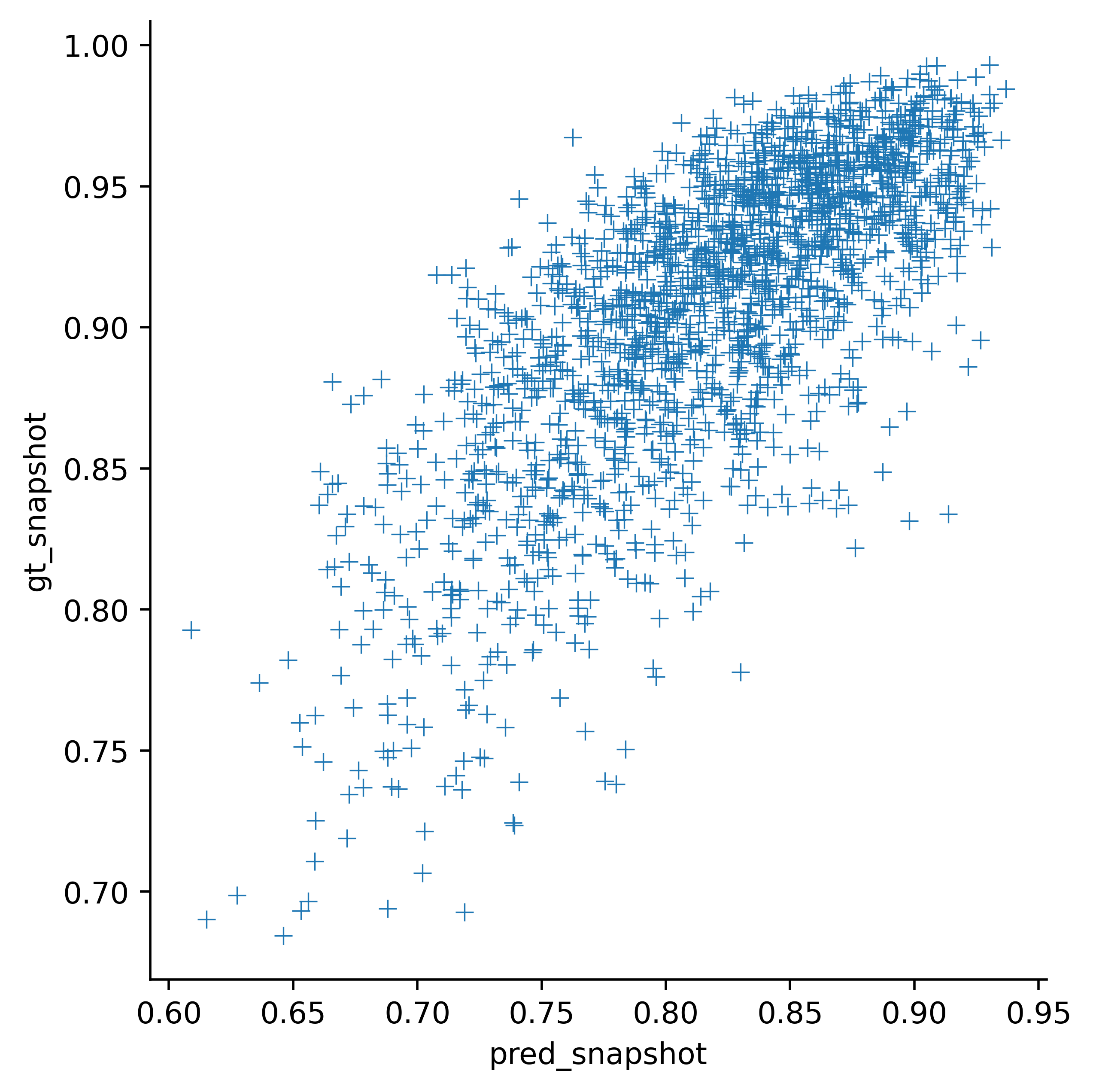}
  \end{center}
  \vspace{-10pt}
  \caption{Ground truth learner performance and predicted learner performance in Simulated-5 validation split}
\vspace{-20pt}
\label{fig:relation_plot}
\end{wrapfigure}

\subsubsection {Comparison between predicted snapshot and true snapshot}

We used Simulated-5, a synthetic dataset, to see the actual discrepancy between true snapshot and predicted snapshot. Fig \ref{fig:relation_plot} shows each learner's ground truth average probability over all questions and corresponding predicted probability in training split of Simulated-5. We observed Pearson correlation of 0.74 and Spearman correlation of 0.72, which is slightly under expectation. This implies improving KT models can result in better diagnosis generation by reducing the gap between predicted snapshot and the ground truth.

\subsubsection{Enhancing trained knowledge tracing model}

Learner interactions produced by our diagnosis generation can slightly improve sequential inference of trained knowledge tracing models. For the experiment, we prepared two different 10-problem sequences : one randomly sampled from the question pool, and the other is sampled by proposed method. We then sampled another random sequence of 10 questions that does not overlap with none of previous two sequences, which is appended to each previous sequence. Then responses were generated for 2000 synthetic learners in the test split using learner knowledge state and IRT model. All generated responses are fed to the trained KT model and last 10 responses are evaluated. For 32 dimension hidden state model, AUC score was improved to 60.52 when using generated diagnosis test, compared to 57.06 when using random test.

% \begin{itemize}
%     \item When optimized with std only, observed best individuals are composed of knowledge concepts with moderate difficulty. This means $C_2$ is correlated to concept difficulty. 
%     \item Correct ratio and standard deviation (discrimination ability) has a unique relationship. std is the highest when correct ratio is in the middle and std is low when correct ratio is too high or too low. 
% \end{itemize}
%% expert results in i-scream edu but not in assistment and ednet

\begin{table}[]
\centering
\resizebox{\columnwidth}{!}{%
\begin{tabular}{ccccc}
\toprule
Problem Id & Related Concept                    & Category             & Curriculum ID & Correct Ratio \\ \midrule
2283       & Real Numbers                       & Number and Operation & 9-1-1-1-3     & 0.39          \\
1318       & Multiplication of Square Roots     & Number and Operation & 9-1-1-2-1     & 0.56          \\
2180       & Multiplication of Square Roots     & Number and Operation & 9-1-1-2-1     & 0.49          \\
389        & Mixed Operations with Square Roots & Number and Operation & 9-1-1-2-2     & 0.33          \\
1563       & Property of Quadratic Function (1) & Functions            & 9-1-4-1-3     & 0.61          \\
278        & Property of Quadratic Function (2) & Functions            & 9-1-4-1-4     & 0.40          \\
1813       & Trigonometric Ratio                & Geometry             & 9-2-1-1-1     & 0.54          \\
804        & Length of Chord                    & Geometry             & 9-2-2-1-1     & 0.58          \\
1932       & Length of Tangent Line to Circle   & Geometry             & 9-2-2-1-2     & 0.29          \\
529        & Inscribed and Central Angle        & Geometry             & 9-2-2-2-1     & 0.50          \\ \bottomrule
\end{tabular}%
}
\caption{\label{tab:QualAnalysis} 9-grader's diagnosis exam generated using genetic algorithm from AIHUBmath Question Pool. Each digit in curriculum id stands for grade, semester, chapter, section, subsection respectively. }
\vspace{-5mm}
\end{table}

\section{Limitation}

We also designed heuristic policy that does utilize prior knowledge on course curriculum, such as subsection information, and problem distribution. Question sheets generated by heuristic policy does not yield competitive $C_1$ and $C_2$, sometimes worse than random baseline. We suspect that predicted snapshot, which we used as a search material, does not contain all the information that defines what makes good diagnostic assessment. Thus our snapshot-based search method is one-way, which means searching over snapshot generates decent diagnostic tests but not all good diagnostic tests give good $C_1$ and $C_2$ on snapshot.  
\section{Related Works}

%% Educational usage / practical use of diagnostic exams
% There has been lots of research on practical effects of diagnostic tests. [...many cites...] 

%% knowledge tracing models - how they were advancing
%% - they are important as a good student simulator! basis for everything
How to assess learner's ability to solve certain problems has been a research interests in education analytic field. Bayesian Knowledge Tracing\cite{corbett1994knowledge} was a monumental research that started using statistical modelings to capture hidden states of learner knowledge. They adopted hidden markov chain to model two discrete knowledge state - know or don't know - and whether a learner answered correctly to a question was used to train it. %% need more properties of BKT 
BKT has a severe drawback that it cannot model multiple question relationships inherently since it only inputs interactions of certain question. To model multiple questions, the same number of models need to be trained.
To alleviate the issue, Learning Factor Analysis\cite{cen2006learning} was introduced that can model several questions. %% 세부 내용
Later as a breakthrough, Deep Knowledge Tracing\cite{piech2015deep} proposed to use Recurrent Neural Network to model successive learner responses from different questions. Question relations can be decoded from DKT using conditional probability that question $j$ was correct given a question $i$ was answered correctly. Also, using this versatile model as a student simulator, one can build a recommender system that gives questions that most boost student performance.
Following success of DKT, many improved model designs appeared including DKVMN\cite{zhang2017dynamic}, SAKT\cite{pandey2019self}, NPA\cite{lee2019creating}, SAINT\cite{choi2020towards}, LANA\cite{zhou2021lana}. They adopted improved modules such as attention, transformer, and student level decoder. 

%% Q matrix design
%% 컨셉별로 분리해서 최적의 진단고사를 내려는 시도의 일종임. 이런 방식의 한계점 소개
Q-matrix design under DINA or DINO model\cite{xu2018empirical} is an attempt for construction of reliable diagnostic assessments. Q-matrix\cite{barnes2005q} is a binary matrix representing question-concept assignments, where $Q_{ij} = 1$ means that underlying concept $j$ is necessary to solve question $i$. Learner's knowledge state can also be represented by a vector of size $K$ when $K$ concepts are present, resulting in $2^K$ possible states for a question requirement, or student state. When we have a student-question response matrix $X$, we can construct a Q-matrix with chosen questions and compute student state using response matrix $X$. Comparing estimated student state and the true state can quantify how well those questions are selected. They did not present how to select a good diagnostic tests. Instead, 3 different Q-matrix were chosen and compared. DINA model is mathematically rigorous but has serious drawbacks. Simply estimating a joint probability distribution of $K$ concepts requires estimation of $2^K$ parameters, making a DINA model infeasible in case of large concept pool. Also their estimation process assumes a \emph{complete} response matrix, which requires every learner solves each question exactly once. It is rarely the case in actual online education settings. In our work, DKT-like models were used which only computes marginal probabilities for each concept state and accepts sparse learner interactions, tackling aforementioned problems of DINA model.

%% diagnostic question paper generation  / difference with our work
\cite{dhavala2020auto} proposed auto generation of diagnostic assessments that well discriminate student performances and can be used to estimate student ability close to ground truth. Authors used 3-parameter Item Response Theory model to estimate ground truth student ability $\theta$ for each question. After diagnostic questions are selected, related interactions are used to train another student IRT model and student ability $\hat{\theta}$ from this partial model is evaluated against $\theta$ above. They claimed policy that incorporate discrimination and student behavior reduced student ability RMSE near 20\%. 

Inspired by \cite{dhavala2020auto}, we reformulated the task so that it does not depend on a specific model like IRT. Learner performance snapshot can be directly compared without any model fitting process. It is also more generic since it can be derived from any models including loopy belief propagation\cite{he2021quizzing}, BKT or even average correct ratio. Consequently, similarity to the whole question pool can be directly incorporated into objective function. For other hard constraints, such as distribution of difficulty levels and syllabus coverage can be forced via search space reduction or filtering of individuals. 
% We chose concept-level search since it has a couple of benefits. Difficulty and knowledge concepts can be decoupled so more systematic assignment of questions are possible. Also as shown in analysis section, discrimination ability of assessment test are closely related to consisting question difficulty, which could be explored easily in concept-level setting.

\section{Conclusion}
In this work, we formulated the task of diagnosis generation as a combinatorial search problem on learner-question snapshot matrix and presented a pipeline that can effectively search for question subsets that maximize proposed criterion. Our proposed formulation is \textbf{generic} in that snapshots can be generated with any known learner performance measures and doesn't even have to be a probability distribution. Also, any search algorithm can be applied to find an optimal subset. Qualitative analysis shows that GA produces fair diagnosis question sheet even though it is searched over predicted snapshot and utilized no prior knowledge on curricula, which asserts its effectiveness in practical education. We expect that improved set modeling or search algorithm can be applied to yield better collection of questions.

%% proir -> prior 수정 (최현성)

\section*{Acknowledgement}

We would like to express special thanks of gratitude to Beomsoo Kang for all the feedback and supports during the write-up and revision process.

%%
%% The acknowledgments section is defined using the "acks" environment
%% (and NOT an unnumbered section). This ensures the proper
%% identification of the section in the article metadata, and the
%% consistent spelling of the heading.

%%
%% The next two lines define the bibliography style to be used, and
%% the bibliography file.
\newpage
\bibliographystyle{ACM-Reference-Format}
\bibliography{main}

%%
%% If your work has an appendix, this is the place to put it.
\newpage
\appendix
\section{Supplementary}

Hyperparameters were searched over validation split, which is part of the training split.
We used the same split scheme for KT model training and genetic algorithm search.

\subsection{KT model Settings and Experiments}
\label{appendix:ktmodel}

\begin{table}[hbt!]
  \begin{tabular}{lcccccc}
    \toprule
    Dataset & $d_q, d_v$ & Batch Size & Learning Rate & Dropout & AUC & ACC \\
    \midrule
    ASSISTment2009 & 256 & 1,024 & 1e-3 & 0.2 & 0.7612 & 0.7319 \\
    ASSISTment2015 & 64 & 1,024 & 1e-3 & 0.2 & 0.7229 & 0.7444 \\
    EdNet & 256 & 1,024 & 1e-3 & 0.2 & 0.7485 & 0.7223 \\
    Simulated-5 & 64 & 1,024 & 1e-3 & 0.2 & 0.7923 & 0.7992 \\
    AIHUBmath & 256 & 1,024 & 1e-3 & 0.2 & 0.8284 & 0.7590 \\
    \bottomrule
  \end{tabular}
  \caption{$d_q$ and $d_v$ stand for hidden dimension of question vector $q$ and learner state $v$}
  \label{tab:freq}
\end{table}

\subsection{Genetic Algorithm Hyperparameters}
\label{appendix:genetic}

\begin{table}[hbt!]
  \begin{tabular}{m{0.2\linewidth}M{1.2cm}M{1.2cm}M{1.2cm}M{1.2cm}M{1.2cm}M{1.2cm}}
    \toprule
    Dataset &  $p_c$ & $p_{m_1}$ & $p_{m_2}$ & $K$ & $N_p$ & $N_g$ \\
    \midrule
    ASSISTment2009 & 0.5 & 0.5 & 0.25 & 25 & 1000 & 5 \\
    ASSISTment2015 & 0.75 & 0.75 & 0.25 & 10 & 1000 & 5 \\
    EdNet & 0.25 & 0.75 & 0.25 & 12 & 1000 & 5 \\
    Simulated-5 & 0.75 & 0.5 & 0.25 & 10 & 1000 & 5 \\
    AIHUBmath (7th) & 0.5 & 0.25 & 0.25 & 10 & 1000 & 5 \\
    AIHUBmath (8th) & 0.5 & 0.75 & 0.25 & 10 & 1000 & 5 \\
    AIHUBmath (9th) & 0.5 & 0.75 & 0.25 & 10 & 1000 & 5 \\
    \bottomrule
  \end{tabular}
    \captionsetup{width=.75\textwidth}
    \caption{$p_c$ is probability of performing crossover operation, and $p_{m_1}$, $p_{m_2}$ stand for probability of choosing an individual to be mutated and probability of each gene to be mutated respectively. $k$, $N_p$, $N_g$ mean number of questions to sample, size of population and number of generations respectively.}
  \label{tab:freq}
\end{table}

\newpage

\subsection{Genetic Algorithm Pseudocode}

\begin{algorithm}
	\caption{Genetic Algorithm} 
	\begin{algorithmic}[1]
	    \State $population \gets \text{A list of some random individuals}$ \Comment{Production}
	    \State $P \gets \text{Size of population}$
    	\State $N_g \gets \text{The number of generations}$
		\For {$generation=1,2,\ldots,N_g$}
		    %% Start Selection
			
			\State $newPopulation \gets \text{empty list}$
			\For {$i = 1,2,\ldots, P$} \Comment{Selection}
			    \State $indiv \gets \text{Randomly select 10\% from population and choose the best individual}$
			    \State $newPopulation.append(indiv)$
			\EndFor
			\State $population \gets \text{newPopulation}$
			
			%% End Selection
			
		    %% Start Crossover

    		\For {$i = 1,2,\ldots, P/2$}     		\Comment{Crossover}
        		\State Select two individuals in $population$
        		\State Generate two offspring by crossover with probability $p_{c}$
        		\State Replace the two individuals with two offspring
    		\EndFor
    		%% End Crossover
    		
    		%% Start Mutation
			\For {$i = 1,2,\ldots, P$} \Comment{Mutation}
			    \State Select each individual in $population$ with probability $p_{m_1}$
			    \State Mutate the genes of selected individual with probability $p_{m_2}$
			\EndFor

			%% End Mutation
			
		\EndFor	
		\State $bestIndividual \gets \text{Select the best individual in population}$
		
		\State \Return $bestIndividual$
	\end{algorithmic} 
\end{algorithm}

\subsection{Qualitative analysis on generated diagnostic tests for 7th and 8th graders}

\begin{table}[hbt!]
\centering
\resizebox{\columnwidth}{!}{%
\begin{tabular}{ccccc}
\toprule
Problem Id & Related Concept                    & Category             & Curriculum ID & Correct Ratio \\ \midrule
1960       & Prime Numbers                       & Number and Operation & 7-1-1-1-1     & 0.55          \\
248       & Common Multiples and Least Common Multiples     & Number and Operation & 7-1-1-2-2     & 0.58          \\
612       & Application of Least Common Multiples     & Number and Operation & 7-1-1-2-2     & 0.66          \\
2651        & Signs of Rational Numbers & Number and Operation & 7-1-1-4-2     & 0.34          \\
2293       & Polynomials & Variable and Expression            & 7-1-2-1-3     & 0.57          \\
243        & Multiplication and Division between Monomials and Numbers & Variable and Expression            & 7-1-2-1-3     & 0.40          \\
998       & Transposition                & Variable and Expression             & 7-1-2-2-2     & 0.42          \\
682        & Position of the point on the coordinate plane                    & Functions             & 7-1-3-1-1     & 0.56          \\
2089       & Perpendicularity of two planes   & Geometry             & 7-2-1-2-1     & 0.60          \\
1351        & Sum of the sizes of internal angles in Polygons        & Geometry             & 7-2-2-1-3     & 0.64          \\ \bottomrule
\end{tabular}%
}
\caption{\label{tab:QualAnalysis} 7-grader's diagnosis exam generated using genetic algorithm from AIHUBmath Question Pool. Each digit in curriculum id stands for grade, semester, chapter, section, subsection respectively. }
\vspace{-7mm}
\end{table}

\begin{table}[hbt!]
\centering
\resizebox{\columnwidth}{!}{%
\begin{tabular}{ccccc}
\toprule
Problem Id & Related Concept                    & Category             & Curriculum ID & Correct Ratio \\ \midrule
1272       & Rational Numbers                       & Number and Operation & 8-1-1-1-1     & 0.47          \\
2809       & Classification of Rational Numbers     & Number and Operation & 8-1-1-1-1     & 0.82          \\
398       & Law of exponents (1) - Addition of exponents     & Variable and Expression & 8-1-1-2-1     & 0.55          \\
1089        & Mixed calculation of Polynomials & Variable and Expression & 8-1-1-3-2     & 0.32          \\
2575       & Solution of first order inequality & Variable and Expression            & 8-1-2-1-3     & 0.67          \\
686        & Application of linear equation system (distance, speed, time)  & Variable and Expression            & 8-1-2-4-1     & 0.57          \\
143       & Relationship between a linear function and a linear equation with two variables                & Functions             & 8-1-3-2-1     & 0.65          \\
2497        & Relationship between a linear equation graph and a linear equation system                    & Functions             & 8-1-3-2-2     & 0.41          \\
1329       & Property of the bisector of the internal angle in triangle.   & Geometry             & 8-2-3-2-1     & 0.46          \\
1552        & Pythagorean Theorem        & Geometry             & 8-2-4-1-1     & 0.56          \\ \bottomrule
\end{tabular}%
}
\caption{\label{tab:QualAnalysis} 8-grader's diagnosis exam generated using genetic algorithm from AIHUBmath Question Pool. Each digit in curriculum id stands for grade, semester, chapter, section, subsection respectively. }
\vspace{-7mm}
\end{table}

\end{document}